\documentclass[aps,floatfix,prd,twocolumn,a4paper,10pt,citeautoscript, longbibliography]{revtex4-1} 

\usepackage{graphicx,color}
\usepackage{dcolumn}
\usepackage{amssymb}
\usepackage{amsmath}
\usepackage{amsfonts}
\usepackage{bm}
\usepackage[colorlinks=true,linkcolor=blue,pagecolor=blue,filecolor=blue,menucolor=blue,urlcolor=blue,citecolor=blue,anchorcolor=blue]{hyperref}

\begin{document} 
\title{Electron cooling mediated by transverse optical phonons in doped SrTiO$_3$} 

\author{Alexander~A.~Zyuzin}
\affiliation{Jo\v{z}ef Stefan Institute, Jamova 39, SI-1000 Ljubljana, Slovenia}

\begin{abstract}
Soft transverse optical (TO) phonons in doped SrTiO$_3$ couple to electrons
through the polarization field, quadratically in the paraelectric phase and,
additionally, linearly in the polar phase due to the onset of spontaneous polarization. We investigate heat flow between a three-dimensional electron gas and soft TO phonons. For two-TO-phonon scattering, the heat flow of hot electrons crosses over from $T^5$ to $T^3$ at the Bloch-Gr\"uneisen temperature. In the polar phase, an additional one-TO-phonon channel emerges, yielding a softer temperature dependence of the heat flow, $T^3$ at low and $T$ at high temperatures, with the spectral weight set by the spontaneous polarization. This channel may dominate at low temperatures and leads to a nonmonotonic temperature dependence of the heat flow in the vicinity of the polar transition.
\end{abstract}
\maketitle

\section{Introduction}
\label{sec:introduction}
Hot electrons in a metal may cool by transferring energy to the lattice. For a coupling linear in the lattice displacement, the energy-transfer rate per unit volume (heat flow) $\mathcal{P}(T_{\rm e},T)$ from electrons at temperature $T_{\rm e}$ to phonons at temperature $T$ exhibits a power-law dependence on temperature. In the standard three-dimensional model, it scales as $T_{\rm e}^{5}-T^{5}$ in the low-temperature regime and becomes linear in $T_{\rm e}-T$ at sufficiently high temperatures \cite{Kaganov1957, Allen1987}. The relevant crossover temperature depends on the phonon spectrum and the electronic phase space. Disorder, reduced dimensionality, and the phonon type may modify the temperature exponents \cite{Reizer1986, Reizer1989, Sergeev_PRL2005}.

In a centrosymmetric crystal, a zone-center TO phonon is odd under inversion so the leading electron-TO-phonon interaction is quadratic in the polarization field $\bm P$,
described by the term $\propto \bm P^2 n_{\rm e}$, where $n_{\rm e}$ is the electron density operator \cite{Ngai1974, Epifanov1981, Epifanov1982}. 
The candidate example is doped SrTiO$_3$ (STO), a paraelectric whose lowest TO mode softens on cooling \cite{Yamada1969, Courtens1993, Chandra2017}. Two-phonon scattering has been used to describe the temperature dependence of the resistivity and thermoelectric coefficients \cite{KumarMaslov2021, NazaryanFeigelman2021}, as well as the dependence of the superconducting transition temperature on density and oxygen isotope \cite{Ruhman_Lee, Marel2019, Volkov2022, Kiseliov_Feigelman2021}; for reviews see \cite{Behnia_review, Fernandes_review, Muller_review}. 

Its consequences for the energy relaxation rate have recently been addressed.
Nonlinear THz spectroscopy of doped STO films found an energy relaxation rate
much slower than momentum relaxation and increasing with temperature
\cite{Kumar2025}, and for two TO phonons the energy relaxation rate was shown theoretically to scale as $T^3$
below and as $T$ above the Bloch-Gr\"uneisen temperature, $T_{\rm BG}$ \cite{CoveyMaslov2026}. 

The interaction may change further in a polar metal. In Ca-doped or
oxygen-isotope-substituted STO, which become polar metals below a transition
temperature $T_{\rm FE}$ \cite{Rischau2017, Wang2019, Rischau2022}, the
polarization acquires a mean value $\bm P_0(T)$. The interaction
$\propto (\bm P_0+\bm P)^2 n_{\rm e}$ then may contain a term linear in the phonon
field, which was shown to produce an anisotropic resistivity and to enhance
superconductivity \cite{ZyuzinZyuzin2022}. Its effect on electron-lattice
heat flow has not been studied.

Here we compute the heat flow between a clean degenerate electron gas and
soft TO phonons, for the two-phonon coupling of the paraelectric phase and
the additional one-phonon coupling of the polar metal. The one-phonon channel may dominate at low temperatures and make the heat flow and the energy-relaxation rate nonmonotonic across $T_{\rm FE}$. This is a thermal fingerprint of the polar phase that complements the anisotropic resistivity of
Ref.~\cite{ZyuzinZyuzin2022}.

\section{Electron-phonon interaction model}
\label{sec:model}
We consider a clean, three-dimensional electron gas coupled to TO phonons. 
The electrons have a spin-degenerate parabolic dispersion, and are described by
$
H_{\rm el}=\sum_{\sigma} \int \frac{d^3k}{(2\pi)^3} \xi_{\bm k} a_{\sigma}^{\dag}({\bm k})a_{\sigma}({\bm k}), 
$
where $\xi_{\bm k}= k^2/2m-\mu$ with $\mu= k_{\rm F}^2/2m$ the chemical potential, $m$ the effective electron mass, $k_{\rm F}$ the Fermi momentum, and $\sigma=\uparrow,\downarrow$ labels the spin projection. The electronic field operator is
$
\Psi_{\sigma}(\bm r)
=
\int \frac{d^3k}{(2\pi)^3} a_{\sigma}(\bm k)
e^{i\bm k\cdot\bm r}
$. Here and below, we set $\hbar=k_{\rm B}=1$.

The phonon degrees of freedom are two degenerate TO branches,
labelled by $\alpha=1,2$. Their Hamiltonian is
\begin{equation}
H_{\rm ph}
=
\sum_{\alpha} \int \frac{d^3q}{(2\pi)^3}
\omega_{\bm q}
\left(b^\dagger_{\bm q\alpha}b_{\bm q\alpha}+\frac{1}{2}\right),
\label{eq:Hph}
\end{equation}
where $b^\dagger_{\bm q\alpha}$ and $b_{\bm q\alpha}$ are the
phonon creation and annihilation operators, respectively.
The TO phonon dispersion is described by
\begin{equation}
\omega_{\bm q}
=
\sqrt{\omega_{\rm TO}^2+s^2q^2}.
\label{eq:spectrum}
\end{equation}
Here $\omega_{\rm TO}$ is the temperature-dependent TO-mode gap and
$s$ is the long-wavelength mode velocity. The gap decreases
on approaching a polar phase and vanishes at the
critical point in the idealized model. We expand the polarization field in the transverse modes \cite{KumarMaslov2021}:
\begin{eqnarray}
\bm P(\bm r)
=
\sum_{\alpha} \int \frac{d^3q}{(2\pi)^3}
\frac{\Omega_0}{\sqrt{4\pi\omega_{\bm q}}}\,
\bm e_{\bm q\alpha}
\left(
b_{\bm q\alpha}e^{i\bm q\cdot\bm r}
+
{\rm h.c.}
\right).
\label{eq:P}
\end{eqnarray}
The parameter $\Omega_0$ specifies the normalization of the
polarization field and is determined by the microscopic ionic
parameters and the convention used for $\bm P(\bm r)$, \cite{KumarMaslov2021}. 
The polarization vectors obey
$\bm e_{\bm q\alpha}\cdot\bm q=0$ and $
\bm e_{\bm q\alpha}\cdot\bm e_{\bm q\alpha'}
=\delta_{\alpha\alpha'}
$.

The electron-phonon interaction near polar phase is taken to be quadratic in
the polarization, \cite{KumarMaslov2021, NazaryanFeigelman2021, ZyuzinZyuzin2022}:
\begin{equation}
H_{\rm ep}
=
\frac{g}{2}\sum_{\sigma}\int d^3\bm r\,
[\bm P^2(\bm r) + 2 {\bm P}_0\cdot {\bm P}(\bm r) ]
\Psi_{\sigma}^\dagger(\bm r)\Psi_{\sigma}(\bm r),
\label{eq:Hep}
\end{equation}
where $g$ is the effective coupling constant and the $P_0^2$ term is absorbed into the definition of the chemical-potential shift. 
At the lowest nonvanishing order in $g$, an electronic
scattering event involves two TO-phonon modes in the paraelectric phase where $P_0=0$. In the polar metal phase, for $T<T_{\rm FE}$ the polarization acquires a mean value $\bm P_0$, so that
 the one-TO-phonon scattering turns on in addition to the two-TO-phonon mode channel, \cite{ZyuzinZyuzin2022}. 

To express the
interaction in momentum space, we define the quadrature
$
\Phi_{\bm q\alpha}
=
\sqrt{\omega_{\bm q}/2}\,
\left(
b_{\bm q\alpha}
+
b^\dagger_{-\bm q\alpha}
\right)
$.
The interaction then takes the form
\begin{align}\nonumber
 H_{\rm ep}
&=
\frac{1}{2}\sum_{\alpha, \alpha_1, \sigma}
\int \frac{d^3q d^3q_1 d^3k}{(2\pi)^9 }
\Gamma^{(2)}_{\alpha\alpha_1}(\bm q,\bm q_1)\times
\\\nonumber
&\times \Phi_{\bm q\alpha}\Phi_{\bm q_1\alpha_1}
a_{\sigma}^\dagger({\bm k+\bm q+\bm q_1})
a_{\sigma}({\bm k})+
\\
&+\sum_{\alpha, \sigma} \int \frac{d^3q d^3k}{(2\pi)^6 } \Gamma^{(1)}_{\alpha}(\bm q)
 \Phi_{\bm q\alpha}
a_{\sigma}^\dagger({\bm k+\bm q})
a_{\sigma}({\bm k}).
\label{eq:Hep-k}
\end{align}
The two-TO-phonon interaction is described by the vertex
$
\Gamma^{(2)}_{\alpha\alpha_1}(\bm q,\bm q_1)
=
(\bm e_{\bm q\alpha}\cdot
\bm e_{\bm q_1\alpha_1}) (g\Omega_0^2)/(2\pi \omega_{\bm q}\omega_{\bm q_1})
$. The quadrature operators contain both phonon creation and
annihilation operators, so the interaction includes processes
involving two-phonon emission or absorption, as well as
processes in which one phonon is absorbed and another is emitted.

\begin{figure}[t!]
    \centering
    \includegraphics[width=6cm]{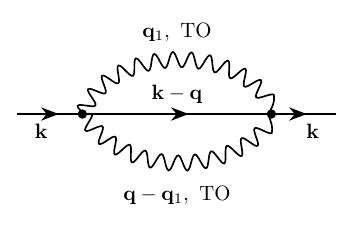} 
     \includegraphics[width=6cm]{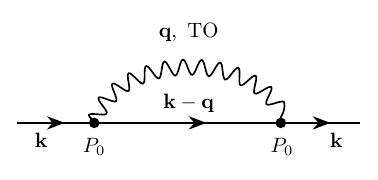} 
    \caption{Top: Diagram describing the electron self-energy due to scattering by the two transverse optical (TO) phonons. The solid and wavy lines stand for the electron and phonon propagators as described below expression (\ref{Collision_integral_main}), while the vertices are described by expression in (\ref{eq:Hep-k}). Bottom: Diagram describing the electron self-energy due to scattering by the one TO-phonon in the polar phase with polarization $P_0$. }
    \label{fig:F1}
\end{figure}

The vertex describing electron scattering by a single TO-phonon reads $
\Gamma^{(1)}_{\alpha}(\bm q)
=
g\Omega_0({\bm P}_0\cdot \bm e_{\bm q\alpha})
/(\sqrt{2\pi} \omega_{\bm q})
$ \cite{ZyuzinZyuzin2022}.
The inverse-frequency factors in electron-phonon vertices Eq.~(\ref{eq:Hep-k}) enhance
the coupling to low-frequency modes. This enhancement makes
the soft-mode regime potentially important for electron
scattering and heat flow. 

\section{Heat flow between electrons and TO phonons}

We consider energy exchange between an electronic subsystem at temperature
$T_{\rm e}$ and a phonon bath at temperature $T$, in degenerate-electron regime $ T_{\rm e},T\ll \mu$. Here we describe energy exchange between two thermal subsystems, characterized by Fermi-Dirac and Bose-Einstein distribution functions at their respective temperatures, rather than the full nonequilibrium distribution that may arise under external driving. 

The kinetic equation for the electron distribution function  is 
$
\partial_t f({\bm k}, \epsilon) =  I_{1}({\bm k},\epsilon)+ I_{2}({\bm k},\epsilon)
$, where the collision integral contains a sum of one- and two-TO phonon interaction processes (labelled $1,2$).

The corresponding contributions to the energy transferred from the electronic subsystem to the phonon subsystem per unit volume and time, $\mathcal{P}= \mathcal{P}_{1}+\mathcal{P}_{2}$, are obtained from the energy-dependent distribution function \cite{Kaganov1957, Allen1987, Sergeev_PRL2005} with the on-shell collision integral calculated along the lines of Refs.~\cite{Altshuler1982, Reizer1989}:
\begin{equation}
\mathcal{P}_{1,2}= -2\int_{-\infty}^{\infty}d\epsilon\,
\epsilon
\int\frac{d^3\bm k}{(2\pi)^3}\,
I_{1,2}(\bm k,\epsilon) \delta(\xi_{\bm k}- \epsilon),
\label{eq:power_collision}
\end{equation}
where  the factor of two accounts
for summation over electron spin. The minus sign is chosen so that $\mathcal{P}>0$ corresponds to the energy flowing from electrons to phonons.

\subsection{Two TO-phonon process}
We first revisit the two-phonon scattering channel \cite{CoveyMaslov2026}, which is present in both
the paraelectric and polar phases. Calculating the two-phonon diagram in Fig.~(\ref{fig:F1}) within the Keldysh representation \cite{Rammer1986, Kamenev2011}, we obtain the self-energy correction to the electronic Keldysh Green function due to scattering by the TO-phonon modes and hence the collision integral:
\begin{widetext}
\begin{align}\nonumber
&I_{2}({\bm k},\epsilon) = 4\int_{-\infty}^{\infty} \frac{d\omega}{2\pi} \int_{-\infty}^{\infty} \frac{d\omega_1}{2\pi}
\int \frac{d^3q}{(2\pi)^3} \frac{d^3q_1}{(2\pi)^3}
\sum_{\alpha\beta}
[\Gamma^{(2)}_{\alpha\beta}({\bm q}_1, {\bm q}- {\bm q}_1)]^2 \times
\\\nonumber
&\times
{\rm Im} G_R({\bm k}- {\bm q},\epsilon-\omega - \omega_1) {\rm Im}  D_R({\bm q}_1,\omega_1) {\rm Im} D_R({\bm q}-{\bm q}_1,\omega)\times
\\
&\times
\bigg\{f(\epsilon)[1-f(\epsilon-\omega-\omega_1)] [1+N(\omega)][1+N(\omega_1)] - [1-f(\epsilon)]f(\epsilon-\omega-\omega_1)
N(\omega) N(\omega_1)  \bigg\}.\label{Collision_integral_main}
\end{align}
\end{widetext}
The Fermi and Bose distribution functions for electrons and phonons are taken at their respective temperatures, $T_{\rm e}$ and $T$: $
f(\epsilon) = [\exp(\epsilon/T_{\rm e}) +1]^{-1}$ and
$N(\omega) = [\exp(\omega/T) -1]^{-1}$.  
The retarded phonon Green function is given by
$
D_{R}({\bm q}, \omega) = \omega_{\bm q}^2/[(\omega+ i\delta)^2-\omega^2_{\bm q}]$ while
 the retarded electron Green function is given by
$
G_{R}({\bm k}, \omega) =(\omega - \xi_{\bm k} + i\delta)^{-1}
$. 
One also uses that $\sum_{\alpha\beta}[\Gamma^{(2)}_{\alpha\beta}({\bm q}, {\bm q}_1) ]^2 =
(g\Omega_0^2/2\pi \omega_{\bm q}\omega_{{\bm q}_1})^2
\mathcal{A}({\bm q},{\bm q}_1),
$
where the polarization sum generates the
geometrical factor
$
\mathcal{A}({\bm q},{\bm q}_1)
=
1+({\bm q}\cdot{\bm q}_1/qq_1)^2,
$
for two transverse branches in an isotropic three-dimensional
system. 

We will focus on the regime when $v_{\rm F}\gg s$ and the temperature is larger than $ms^2$.
In this case the integration over the momentum transfer $q$ is cut by $2 p_{\rm F}$ from above in (\ref{Collision_integral_main}), \cite{KumarMaslov2021}. After rescaling momenta by $T_{\rm e}/s$ ($q$ and $q_1$ are now dimensionless), the heat flow takes the form
\begin{equation}
\mathcal{P}_2 =
 \frac{ \nu T_{\rm e}^5}{\pi^2  E_0 T_{\rm BG}}
\int_0^{T_{\rm BG}/T_{\rm e}}dq\,q\,
F\!\left(q,\frac{T_{\rm e}}{T}\right),
\label{eq:power_scaling}
\end{equation}
where $T_{\rm BG} = 2sp_{\rm F}$ and $\nu= mp_{\rm F}/2\pi^2$ 
is the electronic density of states per spin at the
Fermi energy and 
$
E_0=64\pi^3s^4/(mg^2\Omega_0^4)
$ is an interaction-dependent energy scale introduced in Ref. \cite{KumarMaslov2021}. 
The dimensionless function $F$ contains
the remaining phonon phase-space integral and the polarization dependence:
\begin{align}\nonumber
&F(q,x) =\sum_{a=\pm}\int_0^{\infty}
dq_1 q^2_1\int_{-1}^{1}\frac{d\eta}{4 \varepsilon_{\bm q_1} 
\varepsilon_{{\bm q} - {\bm q}_1} } \mathcal{A}(\bm q_1,\bm q -\bm q_1 )\times
\\\nonumber
&\times
\bigg[
\coth\left(\frac{\varepsilon_{\mathbf{q}_1}+a\varepsilon_{\mathbf{q}-\mathbf{q}_1}}{2}\right)
-\coth\left( x \frac{\varepsilon_{\mathbf{q}_1}+a\varepsilon_{\mathbf{q}-\mathbf{q}_1}}{2} \right)
 \bigg]
 \times
\\
&\times
\bigg[
\coth\left(\frac{ a x \varepsilon_{\mathbf{q}_1}}{2} \right)+ \coth\left(\frac{x \varepsilon_{\mathbf{q}-\mathbf{q}_1}}{2}\right) 
\bigg] 
(\varepsilon_{\mathbf{q}_1} + a\varepsilon_{\mathbf{q}-\mathbf{q}_1})^2,~~
\end{align}
where $\varepsilon_{\mathbf{q}} = \sqrt{\omega_{\rm TO}^2/T_{\rm e}^2 + {\bm q}^2}$ and $\eta = \cos\theta$ with $\theta$ being the angle between vectors $\mathbf{q}$ and $\mathbf{q}_1$. 
The expression for $\mathcal{P}_2$ 
vanishes identically at $T_{\rm e}=T$, as required
by detailed balance. We next consider the soft-mode limit,
$T, T_{\rm e}, T_{\rm BG}\gg \omega_{\rm TO}$, 
in which the TO phonon dispersion may be linearized $
\omega_{\bm q}
\approx sq
$.

Since the integrand in $F$ couples the electron and phonon temperatures, we evaluate the heat flow in distinct asymptotic regimes. For hot electrons, $T_{\rm e}\gg T$, we obtain
\begin{equation}
\mathcal{P}_2 = \begin{cases}
\dfrac{24}{5\pi^2}(9-8\ln2) \zeta(5) \dfrac{\nu T_{\rm e}^5}{ T_{\rm BG}E_0}
, & T_{\rm BG}\gg T_{\rm e}\gg T  \\[1em]
\dfrac{2\zeta(3)}{\pi^2}  \nu \dfrac{T_{\rm BG} T_{\rm e}^3}{E_0}, & T_{\rm e} \gg T_{\rm BG},T.
\end{cases}
\end{equation}
The same leading temperature scalings apply when the electron and phonon temperatures are interchanged, with the different numerical coefficient and sign of the heat flow reversed, provided the corresponding asymptotic conditions are satisfied. 

Near the equilibrium, for $|T_{\rm e}-T|\ll T$, the heat flow is linear in the
temperature difference:
$
\mathcal{P}_2\simeq 19.5 \nu(T_{\rm e}-T)T^4/(T_{\rm BG}E_0)$ at $T\ll T_{\rm BG}, 
$ and 
$\mathcal{P}_2 \simeq 1.5 \nu (T_{\rm e}-T)T^2 T_{\rm BG}/E_0$, at $ T\gg T_{\rm BG}$. The numerical prefactors result from the evaluation of the corresponding dimensionless integral in $F$. These results agree with the corresponding energy-relaxation-rate
scalings obtained in Ref. \cite{CoveyMaslov2026}. 

Finally, when the TO-phonon gap exceeds both temperatures,  $\omega_{\rm TO}> T_{\rm e}, T$, the heat flow due to TO phonons is exponentially suppressed.

\subsection{Single TO-phonon process}
In the polar phase, at second order in $g$ the electron self-energy has an additional contribution to the two-phonon diagram of Fig.~(\ref{fig:F1}), which contains the vertex $\Gamma_{\alpha}^{(1)}(\bm q)$ at both ends and a single phonon line \cite{ZyuzinZyuzin2022}.  
For this one-phonon 
diagram, using the derivations in Ref. \cite{Rammer1986}, 
one obtains the collision integral in the form:
\begin{align}\nonumber
&I_{1}(\bm k,\epsilon)=-4\int_{-\infty}^{\infty} \frac{d\omega}{2\pi} \int \frac{d^3q}{(2\pi)^3}\sum_\alpha\big[\Gamma^{(1)}_\alpha(\bm q)\big]^2 \times \\
&\times
{\rm Im}D_R(\bm q,\omega) {\rm Im}G_R(\bm k-\bm q,\epsilon-\omega)\Big\{f(\epsilon)[1-f(\epsilon-\omega)]\times
\nonumber\\
&\times [1+N(\omega)] -[1-f(\epsilon)]f(\epsilon-\omega)N(\omega)\Big\}.
\label{eq:I1}
\end{align}
Substituting (\ref{eq:I1}) into (\ref{eq:power_collision}), we obtain
\begin{equation}
\mathcal{P}_1=\frac{\nu E_1^2}{T^2_{\rm BG}}\int_{\omega_{\rm TO}}^{ \sqrt{\omega_{\rm TO}^2+T^2_{\rm BG}}}d\omega \omega^2 \left[
\coth \frac{\omega}{2T_{\rm e}} - \coth\frac{\omega}{2T} \right],
\end{equation}
where $E_1^2(T) = (2/3) \nu g^2\Omega_0^2 P_0^2(T)$ and $P_0^2(T) \propto 1- T/T_{\rm FE}$. The one-phonon channel exists only below $T_{\rm FE}$, and we assume that $P_0$ is determined by the lattice temperature $T$. In the soft mode limit $\omega_{\rm TO}\rightarrow 0$, we obtain
\begin{equation}
\mathcal{P}_1 = \begin{cases}
 4\zeta(3) \nu (T_{\rm e}^3-T^3)\dfrac{E_1^2(T)}{T^2_{\rm BG}} 
, & T_{\rm BG}\gg T_{\rm e}, T \\[1em]
 \nu (T_{\rm e}-T)E_1^2(T) , & T_{\rm e}, T\gg T_{\rm BG}.
\end{cases}
\end{equation}
The one-phonon result has therefore the structure of the classical law  \cite{Kaganov1957, Allen1987},  $\propto T_{\rm e}-T$ for $T\gg T_{\rm BG}$, but at low temperatures its exponents are reduced by two compared with the standard deformation-potential result $\propto T_{\rm e}^5-T^5$. The reason is the vertex $\Gamma^{(1)}_\alpha(\bm q)\propto 1/\omega_{\bm q}$ in Eq.~(\ref{eq:Hep-k}), which enhances the coupling to long-wavelength modes and lowers the power of $T$ by $T^2$ relative to a coupling that vanishes at small $q$. Above $T_{\rm BG}$ the momentum transfer is limited by $2p_{\rm F}$ and the phonon occupation is classical. 

Let us compare the phonon channels. 
For $T_{\rm e}\to T$, in both temperature regimes, $T\ll T_{\rm BG}$ and $T\gg T_{\rm BG}$, the ratio has the same form, up to a numerical factor of order unity:
\begin{eqnarray}
\frac{\mathcal{P}_1}{\mathcal{P}_2}\propto\frac{E_0E_1^2(T)}{T_{\rm BG}\,T^{2}}
 \propto \frac{s^3 P_0^2(T) }{\Omega_0^2 T^2}.
\label{eq:ratio}
\end{eqnarray}
The heat flow in the vicinity of the polar phase is nonmonotonic in $T$, a is a feature that the two-phonon channel alone cannot produce. 

\section{Discussion and conclusions}
\label{sec:discussion}

The nonlinear THz experiments on doped STO films \cite{Kumar2025} report a temperature dependence of the electron energy relaxation rate. A comparison of paraelectric and polar phases (e.g., Ca-doped STO) would test the one-phonon channel. In the polar phase, we propose an additional contribution in the electron-phonon heat flow due to a single-TO phonon channel. Using the electronic heat capacity per unit volume $2\pi^2\nu T/3$, the phonon contribution to the energy relaxation rate is $1/\tau_{\rm E}= 1/\tau^{(1)}_{\rm E} + 1/\tau^{(2)}_{\rm E}$ with  $1/\tau^{(1,2)}_{\rm E} = (2\pi^2\nu T/3)^{-1} \partial \mathcal{P}^{(1,2)}/\partial T_{\rm e} |_{T_{\rm e}=T} $. 

In the soft-mode regime, one finds the two-TO-phonon contribution $ 1/\tau^{(2)}_{\rm E} \simeq 3 T^3/(T_{\rm BG}E_0)$
at $T\ll T_{\rm BG}$ and 
$1/\tau^{(2)}_{\rm E} \simeq 0.6 (T_{\rm BG}/E_0)T$ at 
$T\gg T_{\rm BG}$, consistent with Ref. \cite{CoveyMaslov2026}.
The one-TO-phonon contribution gives 
\begin{equation}
\frac{1}{\tau^{(1)}_{\rm E}} = \begin{cases}
 \dfrac{18\zeta(3)}{\pi^2} \left(\dfrac{E_1(T)}{T_{\rm BG}}\right)^2 T
, & T_{\rm BG}\gg T \\[1em]
 \dfrac{3}{2\pi^2}\dfrac{E_1^2(T)}{T} , & T\gg T_{\rm BG}.
\end{cases}
\end{equation}
For example, at temperatures $T>T_{\rm FE}>T_{\rm BG}$, only the two-TO-phonon channel exists and the relaxation rate decreases on cooling. Below $T_{\rm FE}$, the one-TO-phonon process turns on, and the one-phonon contribution to $1/\tau_E$ increases on cooling in the interval $T_{\rm FE}>T>T_{\rm BG}$. A measurement of the energy relaxation in doped STO across $T_{\rm FE}$ may provide a thermal fingerprint of the polar phase, complementary to the anisotropic resistivity of Ref.~\cite{ZyuzinZyuzin2022}.

Our results have several limitations.  The model is a clean, three-dimensional electron gas coupled to two degenerate TO branches. Disorder and phonon damping can change the exponents, as known for the linear coupling \cite{Reizer1986, Reizer1989, Sergeev_PRL2005}. Furthermore, the order parameter $P_0(T)$ is taken in the mean-field form, so critical fluctuations in the immediate vicinity of $T_{\rm FE}$ are not considered. We nevertheless expect the main qualitative prediction to be robust. Namely, below $T_{\rm FE}$ the one-phonon channel, whose weight is fixed by the symmetry-breaking order parameter, $\propto P_0^2(T)$, adds a contribution to the heat flow that is absent in the paraelectric phase.

To conclude, we have calculated the heat flow between a degenerate electron gas and soft transverse optical phonons in the paraelectric and in the polar phases, using doped SrTiO$_3$ as a representative material system. 
The two-TO channel exhibits a crossover in its temperature dependence from $T^5$ to $ T^3$ as temperature increases. In the polar phase at $T<T_{\rm FE}$, the emergence of a finite mean-field polarization activates an additional one-TO-phonon channel, whose contribution has a softer temperature dependence, crossing over from $T^3$ to $T$. Its prefactor is proportional to $1-T/T_{\rm FE}$, reflecting the temperature dependence of the polarization. 

We propose that the interplay between these two channels may lead to a nonmonotonic temperature dependence of the heat flow and the energy-relaxation rate in the vicinity of $T_{\rm FE}$.

\begin{acknowledgments}
The author appreciates E. V. Blinova for strong support. 
We also thank Aalto University where this work was initiated for hospitality. The work was performed within the Marie Sklodowska-Curie Actions programme (GA-101177446) and the Slovenian Research and Innovation Agency (ARIS), contract number 5110-18/2025-5. 
The Pirinem School of Theoretical Physics is warmly acknowledged.
\end{acknowledgments}

\bibliography{Heat_flow_references_arxiv}

\end{document}